
\magnification=1200
\hsize=13.9truecm
\vsize=21.9truecm
\font\ita=cmti10
\overfullrule0pt
\def\ii{\'\i}
\def\ie{\^\i}
\def\de{\delta_\epsilon}
\def\la{\lambda^a}
\def\it12{\int_{\tau_1}^{\tau_2}}
\def\ea{\epsilon^a}
\def\pa{{\cal P}_a}

\def\qc2{\vec q^*_{c_2}}
\hfill PAR-LPTHE-92-18

\hfill ULB-PMIF-92-03
\vskip5.5pc
\centerline{\bf GAUGE INVARIANCE FOR GENERALLY COVARIANT SYSTEMS }

\vskip2.5pc

\centerline{Marc Henneaux $^{(1,2,3,}$\footnote{$^{a)}$}{Ma\ie tre de
recherches au Fonds National de la Recherche Scientifique.}}

\vskip0.9pc

\centerline{Claudio Teitelboim $^{(3,4,5)}$}

\vskip1pc

\centerline{and}
\vskip0.9pc

\centerline{J. David Vergara $^{(2,6)}$}

\vskip2.5pc

\noindent{$^{(1)}$L.P.T.H.E., Universit\'e Pierre et Marie Curie, Tour
16,
1er \'etage}

\noindent{4, Place Jussieu, 75252 Paris Cedex 05, France}
\vskip0.9pc

\noindent{${(2)}$Facult\'e des Sciences, Universit\'e Libre de
Bruxelles,}

\noindent{Campus Plaine C.P. 231, 1050 Bruxelles, Belgium}

\vskip0.9pc

\noindent{$^{(3)}$Centro de Estudios Cient\ii ficos de Santiago,
Casilla 16443}

\noindent{Santiago 9, Chile}
\vskip0.9pc

\noindent $^{(4)}$Facultad de Ciencias, Universidad de Chile, Casilla
653,
\noindent Santiago, Chile.

\vskip0.9pc
\noindent{$^{(5)}$Institute for Advanced Study, School of Natural
Sciences,}

\noindent{Princeton, NJ 08540, USA}
\vskip0.9pc

\noindent{$^{(6)}$Instituto de Ciencias Nucleares, U.N.A.M.}

\noindent{Apto. Postal 70-543, M\'exico D.F., M\'exico}
\vskip5.5pc

\vfill
\eject
\baselineskip1.6pc
\voffset.7truein
\centerline{\bf Abstract}
\vskip2.5pc
Previous analyses on the
gauge invariance of the action for a generally covariant system are
generalized. It is shown that if the action principle is properly
improved, there is as much gauge freedom at the
endpoints for an arbitrary gauge system as there is for a system with
``internal'' gauge symmetries. The key point is to correctly identify
the
boundary conditions for the allowed histories
 and to include the appropriate end-point
contribution in the action. The path integral is then
discussed. It is proved that by employing the improved action,
one can use time-independent canonical gauges even in the case of
generally covariant theories. From the point of view of the action
and the path integral, there is thus no conceptual difference
between general covariance and ``ordinary gauge invariance''. The
discussion is illustrated in the case of the point particle, for which
various canonical gauges are considered.

\vfill
\eject

\centerline {\bf 1. Introduction}

\vskip1.5pc
A generally covariant action is one invariant under reparametrization
of the manifold on which the fields are defined. Important examples
of theories which are generally covariant in their standard
formulation are the theory of gravitation on a closed space and
string theory. When one passes to the hamiltonian formulation of
these theories one finds that there are constraints among the
canonical variables and, morever, the hamiltonian is a linear
combination of the constraints. In Yang-Mills theory, on the other
hand, one also finds constraints, but besides them there is a
non-zero hamiltonian, the total energy. This difference has led
people over the years to regard general covariance as a gauge
symmetry deserving special consideration, since a straightforward
approach would seemingly imply no motion because the hamiltonian
vanishes. In the early days one even spoke of the ``frozen
formalism''.

However the difference hangs from a thin line. A zero-hamiltonian is
a consequence of general covariance only when the canonical
coordinates are taken to be scalars under reparametrization $\tau \to
f(\tau)$ of the time variable. This happens to be the case in the
standard formulation of gravity and string theories, but it is by no
means a theoretical necessity. Indeed, it suffices to perform a $\tau
-$dependent canonical transformation to obtain new variables that are
not scalars and whose hamiltonian does not vanish. Conversely given a
system with a non-vanishing hamiltonian one may always perform a time
dependent canonical transformation that brings the hamiltonian to
zero. In short, the difference between general covariance and
``ordinary gauge symmetries'' is not invariant under time dependent
canonical transformations.

The absence of a clear distinction between general covariance and
ordinary gauge invariance is particularly dramatic in $2+1$
dimensions, where one can reformulate the vacuum Einstein theory as a
Chern-Simons theory with gauge symmetries of the Yang-Mills type [1].
In that case, the changes of coordinates can be expressed as ordinary
internal gauge symmetries modulo trivial gauge transformations
vanishing on-shell.

The purpose of this paper is to reanalyse two
precise issues on which generally covariant systems have been argued
to differ from ordinary gauge systems. The first of these issues
is the invariance of the action under gauge transformations
that do not vanish at the time boundaries. The second -related- issue
is whether  canonical gauge conditions are permissible
in the path integral.  On both issues we find that, if the action
principle
is properly modified by a surface term, one can treat generally
covariant
systems just as ordinary gauge systems of the Yang-Mills type.  Thus
one
can go beyond the limitations previously found in [2], which were
based on the standard form of the action for generally covariant
systems.
The standard form is too restrictive in this sense.

Although we deliberately restrict the analysis of this
paper to precise questions dealing with the transformation properties
of the action and do not deal with issues of interpretation,
 we believe that our results support the view
that generally covariant systems can be quantized
{\ita  \`a la} Yang-Mills. This viewpoint agrees with the results found
in
[3] for 2+1 gravity, as well as with the reduced phase space
discussion given in [4,5].

\vfill
\eject
\vskip1.5pc

\centerline {\bf 2. Gauge invariance of the action}

\vskip1.5pc

\noindent 2.1 Action principle.

\vskip1pc
Consider a gauge system with canonical coordinates $q^i$, $p_i$
($i = 1,...,n$), first class constraints $G_a(q,p)\approx 0$ ($a =
1,...,m$)
and first class Hamiltonian $H_0(q,p)$,

$$
G_a(q,p)\approx 0 \eqno(2.1)$$
$$
[G_a, G_b] = C_{ab}^{\ \ c} (q,p) G_c \eqno(2.2a)$$
$$
[H_0, G_a] = V_a^{\ b} (q,p) G_b \eqno(2.2b)$$
We assume for simplicity that the degrees of freedom and the
constraints are bosonic and that there are no second class
constraints.

Our starting point is the variational principle in the class of paths
$q^i(\tau)$, $p_i(\tau)$, $\lambda^a(\tau)$ taking prescribed values
of a complete set of commuting variables $Q^i(q,p,\tau)$ at the
endpoints $\tau_1$ and $\tau_2$,

$$
Q^i(q(\tau_1),\ p(\tau_1), \tau_1)=Q_1^i \eqno(2.3a)$$

$$
Q^i(q(\tau_2),\ p(\tau_2), \tau_2)=Q_2^i \eqno(2.3b)$$

$$
[Q^i , Q^j] =0 \hbox{ (at equal times)} \eqno(2.3c)$$
These commuting variables are in equal number as the $q$'s
(``completeness'')
 and
 could be for instance the $q$'s themselves ($Q^i = q^i$)
or the $p$'s ($Q^i = p^i$). The action for the variational principle in
which the $Q$'s are fixed at the endpoints is

$$
\eqalign{S[q^i(\tau),p_i(\tau),\la(\tau)]&=
\it12 (p_i \dot q^i - H_0 -\la G_a) d\tau -
B(\tau_2) + B(\tau_1)\cr
&\qquad \hbox{\hskip4.5pc (for paths obeying $(2.3)$)}\cr} \eqno(2.4)$$
 where the phase space function
$B(q,p,\tau)$ is such that
$$
p_i \delta q^i = P_i \delta Q^i + \delta B \eqno(2.5)$$
(for fixed $\tau$). Here, the $P$'s are the momenta conjugate to the
$Q$'s,
$$
[P_i ,  P_j] =0, \ \ \ \ [Q^i , P_j]=\delta^i_j \eqno(2.6)$$
The action $(2.4)$ has an extremum within
the class of paths defined by $(2.3)$ since one finds that when the
equations of motion and the constraints hold,

$$
\delta S = \it12 {d \over d\tau} (p_i \delta q^i) -
[\delta B]_{\tau_1}^{\tau_2} =0 \eqno(2.7)$$
thanks to $(2.5)$ and $\delta Q^i(\tau_1)=\delta Q^i(\tau_2)=0$.

It should be noted that $B$ is determined only up to the addition of a
function of $Q^i$. The addition to $B$ of the function $V(Q^i)$ amounts
to modifying $P_i$ as $P_i \to P_i - \partial V / \partial Q^i.$

\vskip1.5pc

\noindent 2.2 Transversality Conditions
\vskip1pc

The infinitesimal gauge transformations read [6,7,4]

$$
\delta_\epsilon q^i = [q^i , G_a \ea] \eqno(2.8a)$$

$$
\delta_\epsilon p_i = [p_i , G_a \ea] \eqno(2.8b)$$

$$
\de \la  ={\partial \ea \over \partial\tau} +
 [\ea , H_0 + \lambda^b G_b ] + \lambda^c \epsilon^b C_{bc}^{\ \ a} -
\epsilon^b V_b^{\ a}.
\eqno(2.8c)$$
Here, the gauge parameters $\ea (q,p,\tau)$ are arbitrary functions
of $q^i$, $p_i$ and $\tau$. We take $\ea$ inside the Poisson bracket
in $(2.8)$ so that the gauge transformations are canonical
transformations. On the constraint surface $G_a \approx 0$, one can
pull $\ea$ out of the bracket. It follows from (2.8a,b) that the gauge
variation of an arbitrary function $F$ of $q^i$ and $p_i$ is given by

$$
\delta_\epsilon F = [F , G_a \ea] \eqno(2.8d)$$

We shall assume that the functions $Q^i(q,p,\tau)$
and their conjugates $P_i(q,p,\tau)$ are such that the
``transversality condition''

$$
{\partial G_a \over \partial P_i} \hbox{ of  maximum  rank on } G_a
\approx 0
\eqno(2.9)$$
holds. This implies that the constraints can be solved for  $m$ of the
momenta conjugate to $Q^i$, say $P_a$ ($a = 1,...,m$).
 Consequently, no gauge transformation leaves a given set
of $Q$'s invariant (the variables $Q^a$ conjugate to $P_a$ are pure
gauge
and necessarily transform under gauge transformations),
while any set of values for $Q^i$ is compatible with $G_a \approx 0$.

The transversality condition (2.9) is by no means necessary for dealing
with
the path integral in non canonical gauges.  However, it becomes
mandatory if
one wants to write the path integral as a sum over phase space paths
obeying
gauge conditions restricting directly the canonical variables $q^i$,
$p_i$,

$$\chi_a(q,p,\tau) = 0 \eqno(2.10)$$
(``canonical gauges'').  Indeed, in order to reach (2.10), one must
replace
the original boundary conditions (2.3) by gauge related ones that
fulfill
(2.10) at the endpoints.  In operator language, this means that one
must
replace the eigenstates $\vert Q^i\rangle$ of the operators
$\hat{Q}^i$ selected by the boundary conditions (2.3),
by gauge related ones that are annihilated by $\hat{\chi}_a$,

$$\vert Q^i\rangle \rightarrow \hbox{exp}(i\hat{\mu}^a \hat{G}_a)
\vert Q^i
\rangle, \eqno(2.11a)$$

$$ \hat{\chi}_a \hbox{exp}(i\hat{\mu}^a G_a) \vert Q^i> = 0.
\eqno(2.11b)$$
This can be achieved only when the states $\vert Q^i\rangle $ do
transform
 under the  gauge
 transformations. For instance, if the states $\vert Q^i \rangle$
were annihilated by the constraints, one would have exp$(i\hat{\mu}^a
\hat{G}_a)
\vert Q^i \rangle = \vert Q^i \rangle$ and it would be impossible to
reach (2.11b).
[We assume in (2.11) that $[\chi_a , \chi_b] = 0$.  This is
permissible because any set of canonical gauge conditions $\bar{\chi}_a
= 0$
can be replaced by an equivalent one $\chi_a = 0$ on $G_a = 0$
($\bar{\chi}_a
= 0, G_a = 0 \Leftrightarrow \chi_a = 0, G_a = 0$) such that $[\chi_a ,
\chi_b] = 0$].

Furthermore, as we shall see, the path integral in a canonical gauge
involves
only paths that fulfill the constraints everywhere, including the
endpoints.  The transversality condition enables one to assume that
$G_a\approx
0$ holds at $\tau_1$ and $\tau_2$: without (2.9), the conditions $Q^i =
Q^i_1$ or
$Q^i = Q^i_2$ could conflict with $G_a \approx 0$.
Because our ultimate goal is to investigate how to impose canonical
gauges
in the path integral, we shall assume from now on that the
transversality
condition holds.

\vfill

\eject

\vskip1.5pc

\noindent 2.3 Invariance of the action under gauge transformations
vanishing at the endpoints.
\vskip1pc

Since the action $(2.4)$ has been defined only for paths obeying
the boundary conditions $(2.3)$, it is only meaningful, at this
stage, to compute its variation for gauge transformations preserving
$(2.3)$. This forces the gauge parameter $\ea$ in (2.8) to vanish
at the endpoints,

$$
\ea (q(\tau_1), p(\tau_1),\tau_1)=0 \eqno(2.12a)$$
$$
\ea (q(\tau_2), p(\tau_2),\tau_2)=0 \eqno(2.12b)$$

The variation of the action $(2.4)$ under the transformation $(2.8)$
is equal to the endpoint term

$$
\left[ p_i {\partial(\ea G_a) \over \partial p_i} - \ea G_a -
[B, \ea G_a] \right]^{\tau_2}_{\tau_1} \eqno(2.13)$$
and hence is zero thanks to $(2.12)$. Therefore, the action $(2.4)$
is invariant under arbitrary gauge transformations vanishing at the
endpoints.

For later use, it is convenient to rewrite $(2.13)$ as

$$ \left[ P_i {\partial G \over \partial P_i} - G
\right]^{\tau_2}_{\tau_1} \eqno(2.14a)$$
where
$$
G\equiv \ea G_a \eqno(2.14b)$$
is expressed in terms of $Q^i$ and $P_i$. The equation $(2.14a)$
follows from

$$
\delta_\epsilon B \equiv [B, G]= p_i \delta_\epsilon q^i - P_i
\delta_\epsilon Q^i \eqno(2.14c)$$
(see $(2.5)$).

\vskip1.5pc
\noindent 2.4 Invariance of the action under gauge transformations
not vanishing at the endpoints.
\vskip1pc

Because of the transversality condition, the boundary conditions are
not
invariant under gauge transformations.  Thus, if the gauge parameters
$\ea$
do not vanish at the endpoints, the
boundary conditions are modified and the class of paths under
consideration is changed.

Now, the action $(2.4)$ has been defined only for those paths that
fulfill $(2.3)$. Therefore, in order to study its transformation
properties, it is first necessary to extend it off that class of
paths. That is, one must adjust the surface term in (2.4) to the new
boundary
conditions. The observation made in [2], that without an appropriate
boundary term the original action (2.4) is not gauge invariant at the
endpoints is correct, but it is just a reflection of the fact that if
one
insists in fixing the $Q$'s at the endpoints, then there is less gauge
freedom. The new development reported here is that, while this may be
convenient in practice, it is not a necessity of principle.

To analyse the issue of how to extend the action, let us first focus on
a
single gauge transformation with a definite choice of $\ea$. The
boundary conditions obeyed by the new paths read

$$
\bar Q^i (\tau_1) \equiv (Q^i - [Q^i , G]) (\tau_1) = Q^i_1
\eqno(2.15a)$$

$$
\bar Q^i (\tau_2) \equiv (Q^i - [Q^i , G]) (\tau_2) = Q^i_2
\eqno(2.15b)$$
since $\de Q^i = [Q^i , G]$. Thus, in terms of the new paths, it is
$\bar Q^i (q,p,\tau)$ that is kept fixed at the endpoints. According to
the
discussion of \S 2.1 above, the action adopted to the new boundary
conditions must differ from the action
$(2.4)$ by the endpoint term $-D(\tau_2) + D(\tau_1)$, where $D$ is
such that
$$
P_i \delta Q^i =\bar P_i \delta \bar Q^i + \delta (D+M) \eqno(2.16a)$$
with
$$
\bar P_i = P_i - [P_i , G] \eqno(2.16b)$$
In $(2.16a)$, $M$ is an infinitesimal function of $\bar Q^i$
reflecting the ambiguity in the surface term.

A direct calculation yields

$$ D+M = P_i {\partial G \over \partial
P_i} - G \eqno(2.17)$$ The action for the new paths obeying the
boundary conditions $(2.15)$ is thus

$$
S[q^i(\tau), p_i(\tau), \la(\tau)] = \it12 ( p_i \dot q^i -
H_0 - \la G_a) d\tau - \left[ B + P_i {\partial G \over \partial P_i
} - G \right]^{\tau_2}_{\tau_1} \eqno(2.18)$$
(for paths fulfilling $(2.15)$).
We have taken $M=0$. This is the only choice that makes the action
invariant as we now show.

Under the gauge transformation generated by $G$, the class of paths
$(2.3)$ is mapped on the class of paths $(2.15)$. The variation of
the action is equal to

$$ \eqalign{ \de S &\equiv S[q^\prime(\tau), p^\prime(\tau),
\lambda^\prime(\tau)] - S[q(\tau), p(\tau), \lambda(\tau)]\cr & = \it12
(
p^\prime_i \dot q^{\prime i} - H_0^\prime - \lambda^{\prime
a} G^\prime_a ) d\tau -
 \it12 ( p_i \dot q^{i} -
H_0 - \lambda^{a} G_a ) d\tau \cr
& \ - [B^\prime ]^{\tau_2}_{\tau_1} +  [B ]^{\tau_2}_{\tau_1} -
\left[ P_i {\partial G \over \partial P_i} - G
 \right]^{\tau_2}_{\tau_1} \cr & =0 \cr}\eqno(2.19)$$
as it follows from $(2.14)$. In $(2.19)$, we have used the notations
$H_0^\prime = H_0(q^\prime , p^\prime) $, $G_a^\prime = G_a
(q^\prime , p^\prime)$ and $B^\prime = B(q^\prime, p^\prime
, \tau)$. Hence, the extension $(2.18)$ of $S$ possesses the essential
property of making the action invariant under the gauge
transformation generated by $G$, even though $\epsilon^a$ does not
vanish at the endpoints.

What we have done so far concerns a single gauge transformation.
However,
one can repeat the analysis in exactly the same way for all gauge
transformations. One extends in that manner the action to the entire
class of
paths obeying the original boundary conditions or any set of gauge
related
ones. While the original class of paths was determined by $n$
boundary conditions at $\tau_1$ and $\tau_2$, the new class of paths is
determined by fewer independent boundary conditions, namely $n-m$.
 The extended action is gauge invariant and reduces to
(2.4) when one freezes the gauge freedom at the endpoints by fixing
the non gauge invariant coordinates $Q^a$ to $Q^a_1$ or $Q^a_2$, in
which case, the boundary conditions reduce to
 (2.3).

Strictly speaking, the construction applies only to paths fulfilling
the
constraints at the endpoints since the gauge transformations may not be
integrable off the constraint surface.  This, however, is all right
because in a canonical gauge, only the paths fulfilling $G_a = 0$
contribute to the path integral (see below), so that only the action
for
paths lying on the constraint surface is needed.

The action (2.18) depends on the gauge parameters $\ea(\tau_1)$ and
$\ea(\tau_2)$
through $G(\tau_1)$ and $G(\tau_2)$.  By the transversality
condition, these can be expressed in terms of the endpoint values of
the
canonical variables, since the canonical transformation connecting two
admissible sets of boundary conditions is unique. The resulting action
no
longer involves $\ea$.

\vfill
\eject

\vskip1.5pc

\noindent 2.5 Reduced phase space action
\vskip1pc

One may describe the action obtained in the previous section in terms
of the reduced phase space.
The reduced phase space is by definition the quotient of the
constraint surface by the gauge orbits (see for example [4]). Let us
thus consider paths $q^i(\tau)$, $p_i(\tau)$, $\la(\tau)$ that
project to the same path $q^{*\alpha}(\tau)$, $p_\alpha^* (\tau)$ in
the reduced phase space,
$$
S=S[q^{*\alpha}(\tau), p_\alpha^* (\tau) ] , \qquad \alpha = 1, \dots
,n-m \eqno(2.21)$$

We define $Q^{*i}(q,p,\tau_k)$ $(k=1,2)$ by the following conditions

$$
[Q^{*i} , G_a] \approx 0 \eqno(2.22a)$$

$$
Q^{*i} = Q^i + M^i_j (Q^j - Q^j_k ) \eqno(2.22b)$$

$$
{\partial (Q^{*i}) \over \partial (q,p) }\hbox{ of rank } n-m
\eqno(2.22c) $$
and we take $Q^{*i} (q,p,\tau)$ to be solutions of $[Q^{*i} , G_a]
\approx 0$ that interpolate between $Q^{*i} (q,p,\tau_1)$ and
$Q^{*i} (q,p,\tau_2)$. The $Q^{*i}$ are gauge invariant functions
that coincide with $Q^i$ when $Q^j = Q^j_1 (\tau =\tau_1)$ or $Q^j_2
(\tau =\tau_2)$. The $Q^{*i}$ are thus reduced phase space functions.
They commute because the $Q^i$ commute,

$$
[Q^{*i} , Q^{*j} ] \approx 0 \eqno(2.23)$$
On the constraint surface, there are $n-m$ independent functions
among the $Q^{*i}$, which we denote by $Q^{*\alpha}$. One may say
that the conditions $Q^i(\tau_k) = Q^i_k$ (on $G_a \approx 0$) (i)
contain
$m$ gauge conditions; and(ii) fix $n-m$ independent gauge invariant
functions
that commute (i.e., a ``complete set of commuting observables''),
namely, the
$Q^{*\alpha}$.  These $Q^{*\alpha}$ define ``the gauge invariant
content''
of the $Q^i$.

It is clear that the projected reduced phase space paths that occur
in the variational principle diagonalize $Q^{*\alpha} (q^* , p^*,
\tau_1)$ and $Q^{*\alpha} (q^* , p^*, \tau_2)$ at the endpoints.
Because the action is invariant under gauge transformations that do
not vanish at the endpoints, one concludes that, what is really kept
fixed at the endpoints in the new variational principle
is the gauge invariant content of the
$Q^i$ [8].

Note that $Q^{*\alpha} (q^* , p^*,
\tau_1)$ and $Q^{*\alpha} (q^* , p^*, \tau_2)$ may be different gauge
invariant functions even if $Q^i$ does not depend on $\tau$,
$Q^i =Q^i(q,p)$. This is because the equations defining $Q^{*\alpha}$
involve not only $Q^i$, but also the endpoint values $Q^i_1$ (or
$Q^i_2$) of $Q^i$.

One can write explicitly the action $S$ in terms of the reduced phase
space coordinates. Because the $Q^*$ are kept fixed at the endpoints,
$S$ must be of the form

$$
S=\it12 \left[ p_\alpha^* \dot q^{*\alpha} -
H_0(q^*,p^*) \right] d\tau - [B^*]_{\tau_1}^{\tau_2} +
\left[ V^*
(Q^*) \right]_{\tau_1}^{\tau_2} \eqno(2.24)$$
for some function $V$ of the $Q^{*\alpha}$. In $(2.24)$, $B^*$ is the
reduced phase space function such that

$$
p_\alpha^* \delta q^{*\alpha} = P_\alpha^* \delta Q^{*\alpha} +
\delta B^* \eqno(2.25)$$
While the first two terms in $(2.24)$ can be defined within the
reduced phase space without reference to the original variational
principle $(2.3)$-$(2.4)$, the term $V^*(Q^*)$ may involve explicitly
$Q_1^i$ and $Q_2^i$.

Example: consider a single conjugate pair with action

$$
S[q(\tau), p(\tau), \lambda(\tau)] = \it12 (p\dot q
- \lambda G) d\tau \eqno(2.26a)$$

$$
q(\tau_1)=q_1 , \qquad q(\tau_2)=q_2 \eqno(2.26b)$$

$$
G= p - {dV \over dq} \eqno(2.26c)$$
The system is pure gauge so that the reduced phase space reduces to a
single point. The only functional of the reduced phase space paths
are thus the constants.

One may extend the action to arbitrary paths $q(\tau),\ p(\tau),
\lambda(\tau)$ not fulfilling $(2.26b)$ along the lines of the
previous sections. One gets

$$S[q(\tau), p(\tau),\lambda(\tau)] = \it12 (p\dot q
- \lambda G) d\tau +V(q_2) -V(q(\tau_2))+V(q_1) -V(q(\tau_1))
\eqno(2.27)$$
On the constraint surface, the action reduces to the constant

$$
S=V(q_2) -V(q_1) \eqno(2.28)$$
If we had started with the variational principle

$$
\delta S=0, \ \ S[q(\tau),\ p(\tau),
\lambda(\tau)] = \it12 (p\dot q
- \lambda G) d\tau \eqno(2.29a)$$
$$
q(\tau_1)= q^\prime_1 ,\qquad q(\tau_2)=q^\prime_2 \eqno(2.29b)$$
in which the $q$'s are fixed to different values at the endpoints, we
would have obtained a different functional of the reduced phase space
paths, namely

$$
S^\prime=V(q_2^\prime) -V(q_1^\prime) \eqno(2.30)$$
Thus, the action is a functional of the reduced phase space paths
that involves also the boundary data $q_1$ and $q_2$.

\vskip1.5pc
\noindent 2.6 Conclusions
\vskip1pc

By appropriately extending the action for the gauge related paths not
fulfilling the original boundary conditions, one can arrange so that it
is
invariant under arbitrary gauge transformations, and not just those
that
vanish at the endpoints.  The extended action contains a surface term
which
is, in general, different from zero.

\vskip2pc

\centerline{\bf 3. Examples}
\vskip1.5pc

\noindent 3.1.``Internal gauge symmetries'' in the coordinate
representation

\vskip1pc

It has become customary to call a system with constraints that are
linear and homogeneous in the momenta ``system with internal gauge
symmetries''. Yang-Mills systems are of this type. For a system with
internal gauge symmetries, the coordinates $q^i$ transform among
themselves.

Because the constraints are linear, homogeneous in the momenta, one
has
$$
p_i {\partial G_a \over \partial p_i} - G_a =0 \eqno(3.1)$$
Hence, if one specifies the $q$'s at the endpoints ($Q^i = q^i$,
``coordinate
representation'' ), one finds that the $B$ and $D$-terms introduced
above vanish. The action

$$
S[q^i(\tau),\ p_i(\tau),
\lambda^a(\tau)] = \it12 (p_i\dot q^i - H_0 - \la
G_a )d\tau \eqno(3.2)$$
without surface term, is gauge invariant. The absence of the surface
term, due to the existence of a choice of canonical variables that
makes the constraints linear and homogeneous in the momenta, is of
great
practical value. But -- as the analysis of this paper shows --, it is
not to be
taken as signaling a basic conceptual difference between the systems
that
obey (3.1) and those that do not.

[Although (3.1) holds for the Yang-Mills field, it does not hold for
the
Freedman-Townsend model [9]. That model is a theory of a two-form gauge
field
in Minkowski space. The constraints are of the form $A_{ij} q^i q^j +
B_i q^i$. Thus there are systems which do not obey (3.1) but whose
gauge symmetries are ``internal'' in the sense of not coupling
neighbouring spacetime points.]

\vskip1.5pc

\noindent 3.2 Parametrized free particle in one dimension
\vskip1pc

The parametrized free non-relativistic particle in one dimension is
the generally covariant system obtained by including the non
relativistic time $t$ among the dynamical variables. In the
Hamiltonian formalism, the parame\-tri\-zed particle is thus described
by
the original canonical variables $q$ and $p$, as well as by $t$ and
its conjugate momentum $p_t$.

The action for paths obeying the boundary conditions

$$
q(\tau_1)= q_1 \qquad t(\tau_1)=t_1 \eqno(3.3a)$$
$$
q(\tau_2)= q_2 \qquad t(\tau_2)=t_2 \eqno(3.3b)$$
($Q^i\equiv (q, t)$) is given by

$$S[q(\tau), p(\tau), t(\tau), p_t(\tau)] = \it12 (p
\dot q +p_t \dot t - N {\cal H}) d\tau \eqno(3.4a)$$
with
$$
{\cal H} = p_t + {p^2 \over 2m} \eqno(3.4b)$$

The gauge transformations read
$$
\delta q =[q,\epsilon {\cal H}] \approx {p\over m} \epsilon
\eqno(3.5a)$$
$$
\delta t =[t,\epsilon {\cal H}] \approx  \epsilon
\eqno(3.5b)$$
$$
\delta p_t \approx 0 \qquad \delta p\approx 0 \eqno(3.5c)$$
and modify the boundary conditions. The gauge invariant extension of
the action off the class of paths $(3.3)$ is given by

$$
\eqalign{ &S[q(\tau), p(\tau), t(\tau), p_t(\tau), N(\tau)]
 = \it12 (p
\dot q +p_t \dot t - N {\cal H}) d\tau \cr
&\qquad\qquad  -{1\over 2} [t(\tau_2) -t_2]{p^2(\tau_2) \over m}
+{1\over 2} [t(\tau_1) -t_1]{p^2(\tau_1) \over m}\cr}
 \eqno(3.6)$$
in which one fixes ``the gauge invariant content of $q$ and $t$'' at
$\tau_1$ and $\tau_2$,
$$
q^*_{t_1}(\tau_1)\equiv \left( q- {p\over m}t +{p\over m} t_1\right)
(\tau_1) =q_1 \eqno(3.7a)$$
$$
q^*_{t_2}(\tau_2)\equiv \left( q- {p\over m}t +{p\over m} t_2\right)
(\tau_2) =q_2 \eqno(3.7b)$$ The action (3.6) and the boundary
conditions (3.7) reduce respectively to the action (3.4) and the
boundary conditions (3.3) if one imposes the gauge conditions
$t(\tau_1) =
t_1$ and $t(\tau_2) = t_2$ at the endpoints.

[To derive (3.6) and (3.7), we have integrated the gauge
transformations in the explicit case when $\epsilon$ involves only
$\tau$. If $\epsilon$ depends also on $q$, $t$, $p$ and $p_t$, there
are corrections proportional to the constraint to both $(3.6)$ and
$(3.7)$. We shall not write explicitly these corrections here as they
are not needed in what follows (as we have already mentioned, only the
paths
fulfilling ${\cal H}=0$ contribute to the path integral in a canonical
gauge)].

\vskip1.5pc

\noindent 3.3 Relativistic free particle
\vskip1pc

The constraint for the relativistic particle--which is a generally
covariant system--reads

$$
{\cal H} = p^2 + m^2 \approx 0 \eqno(3.8a)$$
with
$$
p^2 \equiv -(p^0)^2 + \sum_{k=1,2,3} (p^k)^2 \eqno(3.8b)$$
The gauge invariant action for paths fulfilling
$$
X^{k*}_{x^0_1}(\tau_1)\equiv \left( x^k - {p^k x^0 \over
p^0} + {p^k
\over p^0} x^0_1 \right) (\tau_1)= x^k_1 \eqno(3.9a)$$

$$
X^{k*}_{x^0_2}(\tau_2)\equiv \left( x^k - {p^k x^0 \over
p^0} + {p^k
v\over p^0} x^0_2 \right) (\tau_2)= x^k_2 \eqno(3.9b)$$
at the endpoints is given by
$$
\eqalign{S [x^\mu(\tau), p_\mu(\tau), N(\tau)]& =\it12 (p_\mu \dot
x^\mu - N{\cal H}) d\tau \cr
& -\left({p^2 -m^2 \over 2p^0} \right)(\tau_2) [x^0(\tau_2)
-x_2^0 ] \cr
& + \left({p^2 -m^2 \over 2p^0} \right)(\tau_1) [x^0(\tau_1)
-x_1^0 ]
\cr} \eqno(3.10)$$
It reduces to

$$
S[x^\mu(\tau), p_\mu(\tau), N(\tau)] =\it12 (p_\mu \dot
x^\mu - N{\cal H}) d\tau \eqno(3.11)$$
for paths which fulfill, in adition to $(3.9)$, the extra condition

$$
x^0 (\tau_1) = x^0_1 \qquad x^0(\tau_2)=x^0_2 \eqno(3.12)$$
(and thus $x^\mu (\tau_1) = x^\mu_1 , \  x^\mu(\tau_2)=x^\mu_2 $).

[There is a factor of $p^0$ in the denominator of $(3.9)$ and
$(3.10)$. On the mass-shell $(3.8)$, these denominators do not vanish
because $p^0=$\break$\pm \sqrt{\vec p^{\ \!\! 2} +m^2} \not= 0$].

\vskip2pc
\centerline{\bf 4. Path integral and canonical gauges}
\centerline{\bf General considerations}
\vskip1.5pc

\noindent 4.1 Summing over equivalence classes of paths
\vskip1pc

The path integral for gauge systems is most accurately described
within the BRST formalism. However, in order to understand the
conceptual points that follow, it is not necessary to resort to that
powerful tool. For this reason, we will adopt here the older point of
view in which the path integral is defined to be the sum over
equivalence classes of gauge related histories of $\exp iS$, with the
appropriate Faddeev-Popov measure [10]. This point of view is correct
when the gauge transformations ``close off shell'', i.e., when the
constraints $G_a$ are chosen in such a way that they form a Lie
algebra, $[G_a, G_b ] =C_{ab}^{\ \ c}G_c$ with $C_{ab}^{\ \
c}=const.$
The BRST treatment, valid in the general case, is given in the
Appendix A.

In order to select a representative in each equivalence class of gauge
related paths, one imposes gauge conditions. If the paths are required
to
fulfill the original boundary conditions (2.3) at the endpoints, the
gauge
freedom is fixed at $\tau_1$ and $\tau_2$.  The residual gauge freedom
is
then given by

$$
\delta q^i = [q^i , \ea G_a ], \ \delta p_i = [p_i , \ea G_a ]
\eqno(4.1a)$$
$$
\delta \la = {\partial \ea \over \partial \tau} +\dots \eqno(4.1b)$$
with
$$
\ea(\tau_1)= \ea(\tau_2)=0, \eqno(4.1c)$$
and is the only freedom to be fixed by the gauge conditions.

The path integral reads [10]

$$
P.I. = \int Dq \ Dp \ D\lambda \ DC \ D\bar C \prod_\tau \delta (\chi)
\exp i S^{\hbox{eff}} \eqno(4.2)$$
with
$$
S^{\hbox{eff}} = \it12 (p_i \dot q^i - H_0 - \la G_a ) d\tau -
[B]^{\tau_2}_{\tau_1} + S^{\hbox{gh}} \eqno(4.3a)$$
$$
S^{\hbox{gh}} = \it12 \bar C \delta_{C} \chi d\tau \eqno(4.3b)$$
In $(4.2)$ the functions $\chi^a$ of the variables $q^i, \ p_i, \
\la$ and their time derivatives are such that the conditions
$$
\chi^a=0 \eqno(4.4)$$
freeze the gauge freedom $(4.1)$. The paths are subject to the
boundary conditions
$$
Q^i(q(\tau_1), p(\tau_1), \tau_1)= Q^i_1 ,\eqno(4.5a)$$
$$
Q^i(q(\tau_2), p(\tau_2), \tau_2)= Q^i_2 ,\eqno(4.5b)$$
and
$$
C(\tau_1)=C(\tau_2)=0, \eqno(4.5c)$$
$$
\bar C(\tau_1)=\bar C(\tau_2)=0, \eqno(4.5d)$$
The multipliers $\la$ are not fixed at the endpoints, that is, they
are summed over. The integration range for $q, \ p$ and $\lambda$ is
the whole real line. The interpretation of $(4.2)$ in terms of the
operator formalism may be found in [4].

The simplest way to freeze the gauge freedom $(4.1)$ is to take the
$\chi$'s to depend on the first time derivatives of the Lagrange
multipliers $\la$ [2],
$$
\chi^a \equiv \dot \la - \psi^a (q,p) =0 \eqno(4.6)$$
(``derivative gauges''). Indeed, the gauge transformations that
preserve $(4.6)$ are characterized by gauge parameters $\ea$ that
obey second order differential equations. This implies, together with
$(4.1c)$ that they vanish (except for ``unfortunate'' choices of
$\psi^a$). The Lorentz gauge in electromagnetism is of the derivative
type.
\vskip1.5pc

\noindent 4.2 Operator insertions- Proper time
\vskip1pc

One can also consider insertions of gauge invariant operators \break
$A[q(\tau), p(\tau),
\lambda(\tau)]$, $$
\de A\approx 0 \eqno(4.7)$$
The corresponding path integral reads
$$
\langle A \rangle = \int Dq \ Dp \ D\lambda \ DC \ D\bar C \prod_\tau
\delta(\chi) \ A \exp i S^{\hbox{eff}} \eqno(4.8)$$
and does not depend on the gauge fixing conditions.

A particular operator insertion that has attracted considerable
attention in the cases of the parametrized non relativistic particle
and the relativistic particle is
$$
A\equiv \theta (T) \eqno(4.9)$$
where $T$ is the proper time associated with the history and $\theta$
is the Heaviside step function. The insertion of $\theta(T)$ in the
path integral restricts the paths contributing to the sum to paths
having positive proper time. This implements causality and can be
generalized to string theory or gravity [2].

The proper time is defined by
$$
T= \it12 N d\tau \eqno(4.10)$$
and is invariant under gauge transformations vanishing at the
endpoints. Since the gauge freedom is frozen at $\tau_1$ and $\tau_2$
in the path integral $(4.8)$, $\theta(T)$ is
a gauge invariant operator and its insertion is permissible.

\vskip1.5pc

\noindent 4.3 Canonical gauge conditions
\vskip1pc

Instead of derivatives gauge conditions, one may implement canonical
gauge conditions of the type
$$
\chi^a (q,p)=0 \eqno(4.11)$$
in the path integral. For this to be possible, one needs to perform a
gauge
transformation at the endpoints in order to enforce $(4.11)$ at
$\tau_1$ and
$\tau_2$. That is, one must replace the original boundary conditions by
gauge related ones that fulfill (4.11).

In the canonical gauge $\chi^a=0$, the sum $(4.8)$ over equivalence
classes of paths becomes

$$
\langle A \rangle = \int Dq \ Dp \ D\lambda \ DC \ D\bar C \prod_\tau
\delta(\chi^a) \ A \exp i S^{\hbox{eff}} \eqno(4.12)$$
where the paths obey instead of (4.5a,b) the boundary conditions
$$
\bar Q^i (q,p,\tau_1)= Q^i_1 \eqno(4.13a)$$
$$
\bar Q^i (q,p,\tau_2)= Q^i_2 \eqno(4.13b)$$
Here, the $\bar Q^i$ are the variables into which the $Q^i$ transform
under the gauge transformation relating the canonical gauges defined
by $Q^i =Q^i_1$ ($Q^i =Q^i_2$) and $\chi^a=0$. The action
in $(4.12)$
contain the boundary term worked out in \S 2.4 and generated by the
gauge
transformation at the endpoints that implements (4.11).

If the gauge invariant function $A$ does not involve the Lagrange
multipliers as one can assume, by using the equations of motion if
necessary,
 the  integral over $\la$ in $(4.12)$ is straightforward
and yields the delta function $\delta(G_a)$. The ghost integral gives
the Faddeev-Popov determinant $\det [\chi^a, G_b]$. Thus $(4.12)$
becomes

$$
\langle A \rangle = \int Dq \ Dp \ \prod_\tau \left(\delta(G_a)
\delta(\chi^b) \det [\chi^a, G_b] \right) \exp iS \eqno(4.14)$$
with $S$ given by $(2.18)$. In the product over $\tau$ in $(4.14)$,
there is one more delta function of the constraints than there are
delta functions of the gauge conditions and Faddeev-Popov determinant
factors. This is because the $\la$ are integrated over at the
endpoints,
while the ghost are kept fixed and the $\chi^a$ are fixing the gauge
only ``inside'' ( the gauge freedom at the endpoints being frozen by
the boundary conditions)\footnote *{Since the equations $\chi^a =0$
define a good gauge slice, the determinant of $[\chi^a, G_b]$ does
not vanish on $G_a\approx 0$. It possesses accordingly a definite
sign on each connected component of the constraint surface. We choose
$\chi^a$ so that the determinant is positive, because the infinite
product
over $\tau$ in $(4.14)$ is otherwise ill-defined.}. Comparison of the
path integral $(4.14)$ with the reduced phase space path integral is
discussed in the appendix B.

The expression $(4.10)$ as such for the proper time is not invariant
under gauge transformations not vanishing at the endpoints. However,
one may extend it off the gauge $t(\tau_1)=t_1$, $t(\tau_2)=t_2$ (non
relativistic particle) or $x^0(\tau_1)=x^0_1$, $x^0(\tau_2)=x^0_2$
(relativistic particle) in such a way that it is invariant. The
relevant expressions are
$$
 T= \it12 N d\tau + t_2 -t(\tau_2) -[t_1 -t(\tau_1)]
\qquad (\hbox{non relativistic particle})\eqno(4.15)$$
and

$$
T= \it12 N d\tau + {x^0_2 -x^0(\tau_2)\over p^0(\tau_2)}
 - {x^0_1 -x^0(\tau_1)\over p^0(\tau_1)}
\quad (\hbox{relativistic particle})\eqno(4.16)$$
These are the expressions that should be used in the path integral in
an arbitrary gauge.

In the ``proper time gauge'' $\dot N=0$, supplemented by the
conditions $t(\tau_2)=t_2$, $t(\tau_1)=t_1$ (or $x^0(\tau_2)=x^0_2$,
$x^0(\tau_1)=x^0_1$) at the endpoints, the proper time becomes
$$
T=N(\tau_2 - \tau_1) \eqno(4.17)$$
and the insertion of $\theta(T)$ in the path integral amounts to
integrating over a positive range for the single integration constant
$N$ ($\dot N =0$) [2].
In a $\tau$-independent canonical gauge, like $t=0$ (non-relativistic
case) or $x^0=0$ (relativistic case), one gets $N=0$,
$t^0(\tau_2)=t^0(\tau_1)=0,$ $x^0(\tau_2)=x^0(\tau_1)=0$ and thus $T$
becomes
$$
T=t_2 -t_1 \qquad (\hbox{non relativistic case}), \eqno(4.18)$$

$$ T={x^0_2\over p^0(\tau_2)} -{x^0_1\over p^0(\tau_1)} \qquad
(\hbox{relativistic case}), \eqno(4.19)$$

\vfill
\eject

\vskip2pc
\centerline{\bf 5. The free non-relativistic particle}
\vskip1.5pc
\noindent 5.1 Gauge $t=0$
\vskip1pc

If one sets $t(\tau_1)=0$ and $t(\tau_2)=0$ in the action $(3.6)$ for
the parame\-tri\-zed non-relativistic particle, one gets the action

$$
S[q(\tau), p(\tau)] = \it12 d\tau (p \dot q) + {1\over2} t_2
{p^2(\tau_2) \over m}-  {1\over2} t_1
{p^2(\tau_1) \over m} \eqno(5.1)$$
on the surface $t=0$, $p_t + {p^2 \over 2m} =0$. The path integral is
thus

$$
P.I.= \int Dq\ Dp \ Dt\ Dp_t \prod_\tau \delta(t) \delta (p_t) \exp iS
\eqno(5.2)$$
(the Faddeev-Popov determinant is unity), with $S$ given by $(5.1)$.
The paths in $(5.2)$ are subject to the boundary conditions $(3.7)$,
which read here ($t=0$)

$$
\left(q +{p\over m} t_1\right) (\tau_1)= q_1 \eqno(5.3a)$$
$$
\left(q +{p\over m} t_2\right) (\tau_2) =q_2 \eqno(5.3b)$$
The integration over $t$ and $p_t$ is direct and yields $1$. Hence,
$(5.2)$ becomes
$$
P.I.= \int Dq\ Dp \exp i\left\{ \it12 d\tau (p\dot q) + {1\over 2m}
\left[t_2 p^2 (\tau_2) -t_1 p^2 (\tau_1)\right] \right\} \eqno(5.4)$$

By making the canonical change of integration variables
$$
Q=q +{p\over m}t_1 +{p\over m}{t_2 -t_1 \over \tau_2 -\tau_1} (\tau
-\tau_1) \eqno(5.5a)$$
$$
P=p \eqno(5.5b)$$
and the time rescaling $\bar \tau ={t_2 -t_1 \over \tau_2 -\tau_1}
(\tau
-\tau_1)+t_1 $, one can recast $(5.4)$ in the form
$$
P.I.= \int DQ\ DP \exp i \left[ \int^{t_2}_{t_1} d\bar \tau \left( P
{dQ \over d\tau} - {P^2 \over 2m} \right) \right] \eqno(5.6a)$$
with
$$
Q(\bar \tau =\tau_1)= q_1 \qquad Q(\bar \tau =\tau_2)= q_2
\eqno(5.6b)$$
This is just the standard path integral for the non-parametrized
particle, known to be equal to
$$
P.I.= [2\pi i(t_2-t_1)]^{-1/2} \exp {i(q_2-q_1)^2 \over 2(t_2 -t_1)},
\eqno(5.7)$$
as it should. Note that $(5.7)$ is a solution of the constraint
equations at the endpoints [11, 4].

One can also compute the quantum average of $\theta(T)$ in the
canonical gauge $t=0$. Since the proper time does not depend on the
paths in the gauge $t=0$ (see$(4.18)$), one can pull $\theta(T)$ out
of the path integral. One gets the causal Green function

$$
 \langle \theta(T) \rangle =\theta (t_2-t_1) [2\pi
i(t_2-t_1)]^{1/2} \exp {i(q_2-q_1)^2 \over 2(t_2 -t_1)}, \eqno(5.8)$$
This is not a solution of the constraint equation at the endpoints
(even though $(5.7)$ is) because the operator being inserted depends
on the endpoints data.

\vskip1.5pc
\noindent 5.2 Gauge $t=p\tau$ --  Time flowing back and forth
\vskip1pc

The path integral may be written in any other canonical gauge by the
same method. The boundary term in the action will take a different
form and a Jacobian different from unity will generically arise. For
example, the gauge

$$
t=p\tau \eqno(5.9)$$
can be reached by a transformation $(3.5)$ with
$$
\epsilon(\tau_2, p(\tau_2))=-t_2 + p(\tau_2) \tau_2 \eqno(5.10a)$$
$$
\epsilon(\tau_1, p(\tau_1))=-t_1 +p( \tau_1) \tau_1 \eqno(5.10b)$$
The Faddeev-Popov determinant is unity and the path integral becomes
$$
\eqalign{ [2\pi  i(t_2-t_1)]^{-1/2} &\exp {i(q_2-q_1)^2 \over 2(t_2
 -t_1)}
=  \int  Dq\ Dp\ Dt\ Dp_t \prod_\tau \delta (t-p\tau) \delta\left(p_t
+{p^2
\over 2m}\right)\cr
& \times\exp i\bigg\{ \it12 (p\dot q + p_t\dot t)d\tau + {1\over
2m}\big[
 p^2(\tau_2) (t_2 -p(\tau_2)\tau_2)\cr &\hskip2pc - p^2(\tau_1) (t_1
-p(\tau_1)\tau_1) \big] \bigg\}\cr}\eqno(5.11)$$
The boundary conditions are
$$
q(\tau_1) - {p^2(\tau_1) \over m} \tau_1 + {p(\tau_1) \over m} t_1
=q_1 \eqno(5.12a)$$
$$
q(\tau_2) - {p^2(\tau_2) \over m} \tau_2 + {p(\tau_2) \over m} t_2
=q_2 \eqno(5.12b)$$
$$
t(\tau_1)=p(\tau_1)\tau_1 \qquad t(\tau_2)=p(\tau_2)\tau_2
\eqno(5.12c)$$

In the gauge $(5.9)$, the histories with constant negative spatial
momentum go back in coordinate time $t$. Thus we see that even for a
non relativistic system time can go backwards.

\vfill
\eject

\vskip2pc
\centerline{\bf 6. The free relativistic particle}
\vskip1.5pc

\noindent 6.1 Gauge $x^0=0$ - Schwinger $\Delta_1$ function [12]
\vskip1pc

The same analysis can be carried out in the case of the relativistic
particle $(3.8)$. In the gauge $x^0=0$, the path integral reduces to
\footnote *{To achieve $[\chi, {\cal H}]= 2 \vert p^0\vert$ (rather
than
$2p^0$), one should really take $\chi = +x^0$ on the sheet $p^0 >0$
and $\chi =-x^0$ on the sheet $p^0 < 0$, e.g. $\chi = \epsilon(p^0)
x^0$.}
$$
P.I.= \int Dx^\mu DP_\mu \prod_\tau \left(\delta(p^2+m^2) \delta(x^0)
2 \vert p^0\vert \right) \exp iS \eqno(6.1)$$
$$
=\int Dx^k Dp_k {1 \over 2\omega (\tau_2)} \exp i S^+
+ \int Dx^k Dp_k {1 \over 2\omega (\tau_2)} \exp i S^-
\eqno(6.2)$$
where $S^+$ (respectively $S^-$) is obtained from $(3.10)$ by setting
$x^0 =0$ and $p^0 =+\omega$ (respectively $-\omega$), with
$$
\omega =\sqrt{\vec p^{\ \! \! 2} +m^2} \eqno(6.3)$$
That is,
$$
S^+ = \it12 p_k \dot x^k d\tau - {m^2 \over \omega(\tau_2)} x^0_2 +
{m^2 \over \omega(\tau_1)} x^0_1 \eqno(6.4a)$$
$$
S^- = \it12 p_k \dot x^k d\tau + {m^2 \over \omega(\tau_2)} x^0_2 -
{m^2 \over \omega(\tau_1)} x^0_1 \eqno(6.4b)$$
The paths in $(6.2)$ are subject to $(3.9)$, i.e.,
$$
\left( x^k + {p^k \omega} x^0_1\right) (\tau_1)= x^k_1 \eqno(6.5a)$$
$$
\left( x^k + {p^k \omega} x^0_2\right) (\tau_2)= x^k_2 \eqno(6.5b)$$
(for $S^+$) and
$$
\left( x^k - {p^k \omega} x^0_1\right) (\tau_1)= x^k_1 \eqno(6.5c)$$
$$
\left( x^k - {p^k \omega} x^0_2\right) (\tau_2)= x^k_2 \eqno(6.5d)$$
(for $S^-$).

To carry out the integral, we make the canonical change of integration
variables
$$
y^k =x^k +{p^k \over \pm \omega} x^0_1 +{p^k \over \pm
\omega}{x_2^0
- x^0_1  \over \tau_2 -\tau_1}(\tau -\tau_1) \eqno(6.6a)$$
$$
P^k =p^k \eqno(6.6b)$$
toghether with the time rescaling $\bar \tau = {x_2^0
- x^0_1  \over \tau_2 -\tau_1}(\tau -\tau_1)+ x^0_1$. This brings the
path integral to the form
$$
\eqalign{ P.I.= &\int Dy DP {1\over 2\omega(\tau_2)} \exp i \left\{
\int_{x_1^0}^{x_2^0} \left[ P_k {dy^k \over d\bar \tau} - \omega
\right] d\bar \tau \right\}\cr & +
\int Dy DP {1\over 2\omega(\tau_2)} \exp i \left\{
\int_{x_1^0}^{x_2^0} \left[ P_k {dy^k \over d\bar \tau} + \omega
\right] d\bar \tau \right\}\cr} \eqno(6.7a)$$
where the paths $y^k(\bar \tau)$ and $P_k(\bar \tau)$ are subject to
$$
y^k (x^0_1) =x^k_1, \qquad y^k (x^0_2) =x^k_2 \eqno(6.7b)$$
By descretizing $(6.7)$ into $N$ time intervals, one finds that the
integral over the intermediate $y$'s yields $N-1$ $\delta$-functions
$\delta( P_k^i - P_k^{i-1})$, leaving one with a single integral over
$P_k$,
$$
\eqalign{P.I. =& \int {dP_k \over 2\omega} \bigg\{ \exp i\left[ P_k
(x^k_2
-x^k_1) -\omega (x^0_2- x^0_1) \right]\cr
& \hskip1.5pc + \exp i\left[ P_k (x^k_2
-x^k_1) +\omega (x^0_2- x^0_1) \right]\bigg\}\cr
&= \int d^4 p \exp [i p_\mu (x^\mu_2 - x^\mu_1)]
\delta(p^2 +m^2)\cr}
\eqno(6.8)$$
This is just Schwinger's $\Delta_1$-function,
$$
P.I.= \Delta_1(x_2, x_1),\eqno(6.9)$$
which is a solution of the homogeneous Klein-Gordon equation.

\vskip3.5pc
\noindent 6.2 Gauge $x^0=0$- Causal propagator
\vskip1pc

One can similarly compute the functional average of $\theta(T)$. The
proper time is given by $(4.19)$ in the gauge $x^0 =0$. For paths
lying on the positive sheet of the mass-shell hyperboloid,
$\theta(T)$ is equal to $\theta\left( x^0_2 - x^0_1
{\omega(\tau_2)\over \omega(\tau_1)}\right)$; for paths lying on the
negative sheet, $\theta(T)$ is equal to $\theta\left( x^0_1 - x^0_2
{\omega(\tau_1)\over \omega(\tau_2)}\right)$. Hence, the path integral
for $\langle \theta(T) \rangle$ reads
$$\eqalign{
 \langle \theta(T) \rangle = \int& Dx^k Dp_k
 {1 \over 2 \omega(\tau_2)}\cr
&\times \left[ \theta\left( x^0_2 - x^0_1
{\omega(\tau_2)\over \omega(\tau_1)}\right) \exp iS^+
 +\theta\left( x^0_1 - x^0_2
{\omega(\tau_1)\over \omega(\tau_2)}\right) \exp i S^- \right].\cr}
\eqno(6.10)$$
By making the same changes of variables as in previous section, the
expression $(6.10)$ becomes
$$
\eqalign{ \langle \theta(T) \rangle =& \theta (x^0_2 -x^0_1) \int
{dP_k \over
2\omega} \exp i\left[P_k(x^k_2 -x^k_1) -\omega (x^0_2
-x^0_1) \right] \cr
& +  \theta (x^0_1 -x^0_2) \int {dP_k \over
2\omega} \exp i\left[P_k(x^k_2 -x^k_1) +\omega (x^0_2 -x^0_1)
\right],\cr} \eqno(6.11)$$
which is the Feynman propagator for the Klein-Gordon equation (up to
the factor $i(2\pi)^{-3}$).

\vskip1.5pc
\noindent 6.3 Light cone gauge $x^+\sim \tau$
\vskip1pc

The calculations of $\langle 1 \rangle$ and $\langle \theta(T)
\rangle$ in the light cone gauge where $x^+$ is proportional to the
time $\tau$ proceed along the same lines. To simplify the derivation,
we take as gauge conditions $$ x^+ = {\tau -\tau_1 \over \tau_2
-\tau_1} (x^+_2 -x^+_1) + x^+_1 \eqno(6.12)$$ so that the boundary
conditions

$$ x^\mu(\tau_2)=x^\mu_2, \qquad x^\mu(\tau_1) =x_1^\mu
\eqno(6.13)$$ fulfill $(6.12)$. The action is then just $\it12
(p_\mu \dot x^\mu - N{\cal H}) d\tau$ without boundary term. The path
integral for $\langle 1 \rangle$ is

$$ \eqalign{ P.I. = \int
Dx^\alpha & Dx^- Dp^\alpha Dp^+ Dp^- \prod_\tau \delta
(p^2 + m^2)
\vert 2p^+\vert \cr
&\times \exp i \it12 d\tau \left[ p_\alpha \dot x^\alpha - p^+
\dot x^- -
p^- \left({ x_2^+ -x_1^+ \over \tau_2 -\tau_1} \right) \right] \cr}
\eqno(6.14)$$
with $\alpha =1,2.$ The paths are subject to the boundary conditions

$$
x^\alpha (\tau_1) =x^\alpha_1 \qquad x^\alpha (\tau_2) =x^\alpha_2
\qquad (\alpha =1,2) \eqno(6.15a)$$
$$
x^- (\tau_1) =x^-_1 \qquad x^- (\tau_2) =x^-_2 \eqno(6.15b)$$

In the product over $\tau$ in $(6.8)$, there is, as before, one more
$\delta$-function of the constraint than there are Faddeev-Popov
factors $\vert 2p^+\vert$.

One can evaluate $(6.14)$ by descretization of the time interval. The
integral over the $x$'s is direct and provides $N-1$
$\delta$-functions $\delta(p_{i+1} -p_i )$. This leaves a single
integral over the momenta $p_\alpha$ and $p^+$. The integral over
$p^-$ is also direct because of the $\delta$-function $\delta(p^2
+m^2)$. The path integral becomes thus

$$
P.I. = \int dp^\alpha {dp^+ \over 2 \vert p^+\vert } \exp
i\left[ p_\alpha
(x_2^\alpha -x_1^\alpha) - p^+ ( x_2^- -x_1^-)
-p^-(x^+_2 - x^+_1)
\right] \eqno(6.16a)$$
with
$$
p^- ={1 \over 2p^+} [(p_\alpha)^2 +m^2] \eqno(6.16b)$$

The expression $(6.16)$ can easily be brought to a more familiar form
by making the change of variables
$$
p^+ ={1\over \sqrt{2}}(p^0 + p^3),\ p^0= \pm \omega(\vec p), \
\omega(\vec p)= \sqrt{ (p_1)^2 + (p_2)^2 + (p_3)^2 +m^2 }
\eqno(6.17)$$
This change of variables maps the negative half line $p^+ < 0$ on the
whole real line for $p^3$, with $p^0 =-\omega(\vec p) < 0$. And
similarly, it maps the positive real line $p^+ > 0$ on the whole real
line $-\infty <p^3 < +\infty $, but this time with $p^0 =\omega (\vec
p) > 0$. Furthermore, $(dp^+) / p^+ = dp^3 /p^0$. Hence, by
separating the $p^+$-integral over $(-\infty, +\infty)$ into an
integral over $(-\infty, 0)$ and an integral over $(0, +\infty)$, one
gets again the standard Fourier representation of $\Delta_1$,

$$
P.I. = \int {dp_k \over 2\omega}  \exp i \vec
p \cdot (\vec x_2
-\vec x_1) \left[ \exp i \omega(\vec p) (x^0_2- x^0_1)
 + \exp -i\omega(\vec p) (x^0_2- x^0_1) \right]
\eqno(6.18)$$

The same steps go through if one inserts $\theta (T)$ with $T$ given
by $(4.19)$ in the path integral.

\vfill
\eject

\vskip2pc

\centerline{\bf 7. Conclusions}

We have shown in this paper that the action for any gauge system may be
made
invariant under gauge transformations not vanishing at the endpoints.
The invariance is achieved by (i) specifying properly what is fixed
at the endpoints; and (ii) including a corresponding boundary term in
the action.

It follows from our analysis that one can use
time-independent canonical gauges (analogous to the Coulomb gauge)
 such as $tr \pi =0$ for gravity on
a compact space, rather than the more familiar $tr \pi = g^{1/2}
\tau$, or $x^+ =0$ for the string, rather than $x^+ =p^+ \tau$.
 The discussion has
been illustrated in the case of the point particle (relativistic
and non relativistic), for which
solutions of the homogeneous wave equation and propagators (causal
Green
functions) have been computed in canonical gauges.

Our analysis assumes throughout that the transversality condition holds
at the endpoints. Locally in phase space, this is not a restriction.
 Indeed, one can always replace locally the constraints by equivalent
ones that are some of the momenta in a new canonical system of
coordinates.
The corresponding $Q$'s are clearly not left invariant by the gauge
transformations and fulfill accordingly the transversality conditions.
However, there may be obstructions to the global existence of such
$Q$'s.  In that case, the transversality condition would necessarily
fail at
"singular" points.  The question depends on the form of the
constraints. The
eventual failure of the transversality condition
is by no means characteristic of generally covariant systems, since we
have given examples of such systems for which the transversality
condition hold globally on the constraint surface (parametrized
systems,
free relativistic massive particle).

Finally, we emphasize that the issues discussed in this paper are
more conceptual than practical. Even though it is permissible to
adopt, in principle, time-independent gauge for generally covariant
systems, such gauges require the explicit integration of the gauge
transformations generated by the constraints in order to adjust the
boundary conditions. This is usually not tractable.

\vskip2pc

\centerline{\bf Acknowledgments}

This work has been partly supported by a research contract with the
Commission of the European Communities, by grants 0862/91, 0867/91 of
FONDECYT
(Chile) and by institutional support provided by SAREC (Sweden) and
Empresas Copec (Chile) to the Centro de Estudios de Santiago. One of
us (M.H.) is grateful to the members of LPTHE-Paris 6 for their
kind hospitality while this work was being brought to completion.

\vfill
\eject

\noindent Appendix A: Reaching a canonical gauge in the BRST path
integral
\vskip1.5pc

The BRST path integral that replaces $(4.2)-(4.3)$ in the general
case of an arbitrary gauge system with closed or open algebra is
[4,13,14]

$$
P.I.= \int Dq\ Dp\ D\lambda\ Db\ D\eta\ D\bar C \ D{\cal P}\ D\rho
\exp iS_K \eqno(A.1)$$

$$
S_K= \it12 d\tau (\dot q^i p_i +\dot \eta^a \pa + \dot \la b_a +x \bar
 C_a \rho^a -H - [K, \Omega] ) - [B]^{\tau_2}_{\tau_1}
\eqno(A.2)$$
where the sum is extended over all paths that obey the boundary
conditions

$$
Q^i (Q(\tau_2), p(\tau_2), \tau_2)= Q^i_2 \eqno(A.3a)$$
$$
Q^i (Q(\tau_1), p(\tau_1), \tau_1)= Q^i_1 \eqno(A.3b)$$
$$
\eta^a(\tau_1)=\eta^a(\tau_2)=0 \eqno(A.3c)$$
$$
\bar C_a(\tau_1)=\bar C_a(\tau_2)=0 \eqno(A.3d)$$
$$
b_a(\tau_1)=b_a(\tau_2)=0 \eqno(A.3e)$$
The path integral $(A.1)$-$(A.2)$ does not depend on the choice of
$K$ [6, 13, 14, 4]. By taking $K$ to be of the form

$$
K= -\pa \la + i \bar C_a \psi^a(q,p) \eqno(A.4)$$
one may perform the integral over $b_a$. This yields the expression
$(4.2)$ with, however, higher order ghost vertices in $S^{\hbox{gh}}$
when
the gauge algebra does not close (and $\eta^a$ is replaced by $C^a$).

In order to implement the canonical gauge $\chi^a(q,p)=0$ $ (\delta Q^i
\not=0)$, we first perform the canonical change of variables

$$
q^{\prime i} =q^i +[q^i, C], \qquad  p^{\prime}_i =p_i
+[p_i, C]
\eqno(A.4a)$$

$$
\eta^{\prime a} =\eta^a +[\eta^a, C], \qquad  \pa^{\prime
} =\pa +[\pa, C]
\eqno(A.4b)$$

$$
\lambda^{\prime a} =\la +[\la, C], \qquad  b^{\prime}_a =b_a
+[b_a, C]
\eqno(A.4c)$$

$$ \bar C^{\prime }_a =\bar C_a +[\bar C_a, C], \qquad
\rho^{\prime}_a =\rho_a +[\rho_a, C] \eqno(A.4d)$$
where $C$ is the BRST-exact function

$$
\eqalignno{ C & = [ -\ea(q,p,\tau) \pa, \Omega] &(A.5a)\cr
& =\ea G_a + \hbox{``higher order terms in }\pa \hbox{ and }\eta^a
\hbox{''}&(A.5b)\cr } $$
The gauge parameters $\ea$ in $(4.5)$ are chosen so that the
modification of $q^i$ and $p_i$ at the endpoints, which reduces to

$$
\eqalignno{ q^{\prime i}& =q^i +[q^i, \ea G_a] &(A.6a) \cr
  p^{\prime}_i & =p_i +[p_i, \ea G_a] &
(A.6b) \cr}$$
because $\eta^a(\tau_1)=\eta^a(\tau_2)=0$, makes the canonical gauge
condition hold at $\tau_1$ and $\tau_2$.

The change of variables $(A.4)$ has the following effect: (i) it
induces in the action the surface term discussed in the text because
the ghosts vanish at $\tau_1$ and $\tau_2$; and (ii) it replaces $K$
by $[H + [K,\Omega], \ea \pa ] -(\partial \ea / \partial \tau) \pa$.
But this second change can be discarded since the path integral
 does not depend on $K$.

The next step in order to reach the canonical gauge $\chi_a =0$ is to
take $K$ to be of the form

$$
K_\epsilon = -\pa \la + {i \over \epsilon} \bar C_a \chi^a
\eqno(A.7)$$
Making the rescaling of variables [6]
$$
b_a = \epsilon B_a, \qquad \bar C_a = \epsilon \bar {\cal C}_a
\eqno(A.8)$$
whose Jacobian is unity, taking the limit $\epsilon \to 0$ (the path
integral does not depend on $\epsilon$), and integrating over $\pa ,
\ \rho^a, \ \la, B_a , \ \bar {\cal C}_a$ and $\eta^a$, one finally
gets the desired path integral expression used in the text

$$
P.I. = \int Dq\ Dp\ \prod_\tau \left( \delta(G_a) \delta(\chi^a) \det
[\chi^a , G_b] \right) \exp iS \eqno(A.9)$$
with $S$ given by $(2.18)$.

The same derivation can be repeated for the quantum averages of BRST
invariant operators (which may be assumed to depend only on the
variables $q, \ p, \eta$ and ${\cal P}$ of the ``minimal sector'').
\vskip2pc

\noindent Appendix B Comparison with the reduced phase space path
integral
\vskip1.5pc

The path integral $(A.1)-(A.3)$ is physically equivalent to the
reduced phase space path integral, but differs from it by endpoint
wave function factors whose explicit form depends on the
representation of the constraint surface [4].

The reduced phase space path integral is invariant under the
redefinitions $G_a \to a_a^b G_b$ of the constraints, while the path
integral $(A.9)$ is not since there is in $(A.9)$ one more $\delta
(G_a)$ than there are Faddeev-Popov determinant factors.

The reduced phase space path integral can also be cast in the form
$(A.9)$,

$$
 \int Dq\ Dp\ \prod_\tau \left( \delta(G_a) \delta(\chi^a) \det
[\chi^a , G_b] \right) \exp i\left[ \it12 (p_i \dot q^i - H_0) d\tau
+ \hbox{bound. term} \right] \eqno(B.1)$$
but the precise meaning of the ``measure'' $\prod_\tau \left(
\delta(G_a) \delta(\chi^a) \det [\chi^a , G_b] \right)$ is now
different. In $(B.1)$, this measure stands for $Dq^* \ Dp^*$ where
$q^{*\alpha}$ and $p^*_\alpha$ form a complete set of reduced phase
space canonical coordinates [One has $dq^{*\alpha} dp_\alpha^* = dq^i
dp_i \delta(\chi^a) \delta (G_a) \det [\chi^a , G_b]$ but in $Dq^*
Dp^*$, there is one ``unmatched'' $dp^*_\alpha$ in, say the
$q^*$-representation. Accordingly, the numbers of $\delta(\chi^a), \
\delta (G_a)$ and $ \det [\chi^a , G_b]$ in $(A.9)$ and $(B.1)$ are
different].

If $Q^{*\alpha}(\tau_1)$ and $Q^{*\alpha}(\tau_2)$ are the reduced
phase space variables kept fixed in the reduced phase space
variational principle, the reduced phase space path integral is equal
to

$$
(Q^{*\alpha}_2 \vert \exp -iH_0 (\tau_2 -\tau_1) \vert Q^{*\alpha}_1)
\eqno(B.2)$$
where $\hat Q^{*\alpha}(\tau_2) \vert Q^{*\alpha}_2) = Q^{*\alpha}_2
\vert Q^{*\alpha}_2)$ and $\hat Q^{*\alpha}(\tau_1) \vert
Q^{*\alpha}_1 = Q^{*\alpha}_1 \vert Q^{*\alpha}_1)$ (note that
$\hat Q^{*\alpha}(\tau_1)$ and $\hat Q^{*\alpha}(\tau_2)$ may be
different
operators). In $(B.2)$, $(\ \vert\ )$ stands for the scalar product
of the reduced phase space quantization. The states $\vert
Q^{*\alpha}_1)$ and $\vert Q^{*\alpha}_2)$ are complete sets of
physical states. Let $\vert Q^{*\alpha}_1 \rangle$ and $\vert
Q^{*\alpha}_2\rangle $ be the corresponding states of the Dirac
quantization, solutions of the constraint equations

$$
\hat G_a \vert Q^{*\alpha}_1 \rangle =0, \qquad \hat G_a \vert
Q^{*\alpha}_2\rangle =0 \eqno(B.3) $$
(these states depend on the form of $\hat G_a$ since the physical
state condition $(B.3)$ does). One may show that the path integral
$(A.1)-(A.3)$ is equal to

$$
\int dQ^{*\alpha}_1 dQ^{*\alpha}_2 \langle Q^{i}_2 \vert
Q^{*\alpha}_2 \rangle ( Q^{*\alpha}_2 \vert \exp -iH_0(\tau_2
-\tau_1) \vert Q^{*\alpha}_1)\langle Q^{*\alpha}_1 \vert Q^{i}_1
\rangle \eqno(B.4)$$
where $\vert Q^{i}_1 \rangle$ and $\vert Q^{i}_2 \rangle$ are
respectively the eigenstates of $\hat Q^{i}(\tau_1)$ and $\hat
Q^{i}(\tau_2)$ with eigenvalues $Q^i_1$ and $Q^i_2$ [4].

The difference between the reduced phase space path integral and the
path integral analyzed in the text is of the same type as--and
possesses
no more significance than--the difference between the kernels of the
same operator in two different representations.

\vfill
\eject

\vskip2.5pc

\centerline{References}
\vskip1.5pc

\item{[1]} E. Witten, Nucl. Phys. B311 (1988) 46.

\item{[2]} C. Teitelboim, Phys.Rev. D25 (1982) 3159.

\item{[3]} S. Carlip, Phys. Rev. D42 (1990) 2647.

\item{[4]} M. Henneaux and C. Teitelboim, ``Quantization of Gauge
Systems'',
 Princeton University Press (Princeton 1992).

\item{[5]} C. Rovelli, Phys. Rev. D43 (1991) 442.

\item{[6]} E.S. Fradkin and G.A. Vilkovisky, CERN Report
TH-2332(1977).

\item{[7]} C. Teitelboim, Phys. Rev. Lett. 38 (1977) 1106.

\item{[8]} J.A. Wheeler, in ``Relativity, Groups and Topology'', C. De
Witt
and B.S. De Witt eds, Gordon and Breach (New York 1964);
 C.Teitelboim, J. Math. Phys. 25 (1984) 1093.

\item{[9]} D.Z. Freedman and P.K. Townsend, Nucl. Phys.
B177 (1981) 282.

\item{[10]} L.D. Faddeev and V.N. Popov, Phys. Lett. B25 (1967) 30.

\item{[11]} J.J. Halliwell and J.B. Hartle, Phys. Rev. D43 (1991)
1170.

\item{[12]} M. Henneaux and J.D. Vergara, ``BRST Formalism and Gauge
Invariant Operators: the Example of the Free Relativistic Particle'',
to appear in the Proceedings of the First International A.D.
Sakharov Conference on Physics (Moscow, May 1991).

\item{[13]}  I.A. Batalin and G.A. Vilkovisky, Phys. Lett. B69 (1977)
309;
E.S. Fradkin and T.E. Fradkina, Phys. Lett. B72 (1978) 343.

\item{[14]} M. Henneaux, Phys. Rep. 129 (1985) 1.

\bye